\documentstyle[12pt,fleqn,twocolumn]{article}
\begin{document}

\noindent
{\bf Conversion of neutron stars to strange stars as an origin
           of $\gamma$-ray bursts}

\vspace{5.0mm}
\noindent K.S. Cheng$^{1}$ \& Z.G. Dai$^{2}$

\noindent
$^{1}${\em Department of Physics, University of Hong Kong, Hong Kong}

\noindent
$^{2}${\em Department of Astronomy, Nanjing University, Nanjing 210093,
China}

\newcommand{\be}{\begin{equation}}
\newcommand{\ee}{\end{equation}}
\newcommand{\pn}{\acute {\rm n}}
\newcommand{\me}{\acute {\rm e}}
\newcommand{\ma}{\acute {\rm a}}

\parindent=5mm
\vspace{5.0mm}
{\small
\noindent
{\bf The recent observational results$^{1}$ of the BASTE detector
strongly suggest that the sources of weak $\gamma$-ray bursts are
at cosmological distances$^{2-4}$. A number of theoretical models
for explaining these sources have also been proposed$^{5}$.
Here we argue that conversion of neutron stars to strange stars
can be another possible origin
of $\gamma$-ray bursts. The converting stars may be
neutron stars in the binaries with low-mass companions.
We show that the energy released per conversion event satisfies the
requirements of cosmological bursts, and the Lorentz factor of the
resultant expanding fireball can exceed $5\times 10^{3}$
because the strange star has very low baryon contamination.
The model burst rate is also consistent with the observed one.
Our model may provide an explanation why the binary millisecond pulsars
having accreted matter with over $0.5M_{\odot}$ seem to have
same low magnetic fields$^{6}$. }

When nucleon matter is squeezed
to a sufficiently high density, it turns into uniform two-flavor quark
(u and d) matter. But the quark matter
is unstable, and subsequently converts to three-flavor (strange)
quark matter, which is due to the fact that
strange matter may be more stable than nucleon matter$^{7}$.
The properties of strange stars have been studied$^{8,9}$.
However, the existence of strange stars
is doubtful. First, the glitch behaviors of pulsars are usually
well described by the neutron-superfluid vortex creep theory (for
a review see Ref.[10]), but in current strange-star models
one hardly explains the observed pulsar glitches. Second,
the conversion of a neutron star to a strange star
requires the formation of a strange-matter seed in the star, which
is produced through the deconfinement of neutron matter at
a sufficiently high density much larger than the central density
of the $1.4M_{\odot}$ star with a rather stiff equation of state$^{11,12}$.
For these two reasons, we assume that the
pulsars except some millisecond ones are ordinary neutron stars.

It is thought that the density for deconfinement of neutron matter
with an intermediately stiff (or stiff) equation state to two-flavor
quark matter is near $8\rho_{0}$ ($\rho_{0}$ is the nuclear
density)$^{11,12}$. For a soft equation of state, the deconfinement
density is lower. Here we assume that the equations of
state in neutron stars are intermediately stiff or stiff. This is
because soft equations of state at high densities are ruled out
by the postglitch recovery in four pulsars$^{13}$.
More detailed analyses of the postglitch curves of the Crab and Vela
pulsars also draw similar conclusions$^{14,15}$.
In addition, the soft equation of state like kaon condensation
seems not to occur in stable neutron stars$^{16}$.
The neutron stars with $1.4M_{\odot}$ based on the modern equations of
state$^{17}$ named UV14+UVII, AV14+UVII and UV14+TNI
must accrete matter of $\sim 0.6M_{\odot}$, $0.5M_{\odot}$ and
$0.4M_{\odot}$ in order that their central densities
reach the deconfinement density. Once this condition
satisfies, strange-matter seeds are formed in the interiors of the stars.

After a strange-matter seed is formed, the strange matter will begin to
swallow the neutron matter in the surroundings.
While it has been proposed$^{18}$ that the combustion corresponds to
the slow mode, subsequent work$^{19}$ shows
that this mode appears to be hydrodynamically unstable.
Thus the conversion of neutron matter
should proceed in a detonation mode.  The total kinetic reaction
of the detonation mode has two stages: the formation of
two-flavor quark matter, and the weak decays that form strange matter.
Since the second process enhances the thermal energy at
the expense of the chemical energy of two-flavor quark matter$^{20}$,
the temperature in the star's interior will increase more than 10MeV.
In addition, the timescale$^{19}$ for the conversion
of a neutron star to a strange star is smaller than 1s.

The resulting strange star$^{21}$ has a thin crust with mass $\sim 2\times
10^{-5}M_{\odot}$ and thickness $\sim 150$m. But the nuclei in
this crust may decompose into nucleons because of the
internal temperature of $10^{11}$K.
Approximating strange matter by a free Fermi
gas, we obtain the total thermal energy of the star,
$E_{{\rm th}}\sim 5\times 10^{51}\,{\rm ergs}\,(\rho/\rho_{0})^{2/3}
R_{6}^{3}T_{11}^{2}\,$, where $\rho$ is the average mass density,
$R_{6}$ the stellar radius in units of $10^{6}$cm,
and $T_{11}$ the temperature in units of $10^{11}$K.
Adopting $\rho=8\rho_{0}$, $R_{6}=1$, and $T_{11}=1.5$, we have
$E_{{\rm th}}\sim 5\times 10^{52}\,{\rm ergs}$.

The star will cool by the emission of neutrinos and antineutrinos.
Because of the huge neutrino number density the neutrino
pair annihilation process $\nu \bar{\nu}\rightarrow e^{+}e^{-}$ operates in
the region close the strange star surface.
The total energy$^{22}$ deposited due to this process is
$E_{1}\sim 2\times 10^{48}\,\,{\rm ergs}\, (T_{0}/10^{11}{\rm K})^{4}
\sim 10^{49}\,{\rm ergs}$ (where $T_{0}$ is the initial temperature).
The timescale for deposition is of the order of 1s.
On the other hand, the processes for $n+\nu_{e}\rightarrow p+e^{-}$ and
$p+\bar{\nu_{e}}\rightarrow n+e^{+}$ play an important role in the
energy deposition. The integrated neutrino
optical depth$^{23}$ due to these processes is
$\tau \sim 10^{-4}\rho_{9}^{4/3}T_{11}^{2}$
(where $\rho_{9}$ is the crust density in units of $10^{9}{\rm g}\,{\rm
cm}^{-3}$). So the deposition energy is estimated by
$E_{2}\sim E_{{\rm th}}(1-e^{-\tau}) \sim 2\times 10^{52}\,{\rm ergs}$.
Here we have used the neutron-drip density, and
have assumed that the thermal energy of the star is wholly lost in
neutrinos. The process, $\gamma \gamma\leftrightarrow e^{+}e^{-}$,
inevitably leads to creation of a fireball.
However the fireball must be contaminated by the baryons in
the thin crust of the strange star. Define
$\eta=E_{0}/M_{0}c^{2}$, where $E_{0}=E_{1}+E_{2}$ is the initial radiation
energy produced ($e^{+}e^{-}$, $\gamma$) and
$M_{0}$ is the conserved rest mass of baryons with which the fireball is
loaded. Since the amount of the baryons contaminating the fireball cannot
exceed the mass of the thin crust, we have $\eta\ge 5\times 10^{3}$.
The fireball will expand outward. The expanding
shell (having a relativistic factor $\Gamma \sim \eta$) interacts with
the surrounding interstellar medium and its kinetic energy is finally
radiated through non-thermal processes in shocks$^{24}$.

What mechanism results in conversion of neutron stars? Here we propose
that accretion in binaries with low-mass companions can lead to
the conversion. We assume that at the beginning of
accretion the masses of neutron stars are $1.4M_{\odot}$.
It has been shown$^{6}$ that the amounts
of matter accreted by
the 18 radio pulsars in these binary systems exceed $0.5M_\odot$.
If this is true, some of the millisecond pulsars
may be strange stars. By assuming that the number of
galaxies at cosmological distances is $\cal N$, and
the number$^{25}$ of the low-mass X-ray binaries with neutron stars
accreting at a high rate near the Eddington limit is
$N_{\rm B}$, we have the burst rate
\noindent
\begin{eqnarray}
{\cal R}  \sim  \frac{{\cal N}N_{\rm B}}{\Delta M/\dot {M}} \sim
10\,{\rm day}^{-1}\left(\frac{{\cal N}}{10^{10}}\right) \nonumber  \\
\times \left(\frac{N_{B}}
{10}\right)\left(\frac{\Delta M}{0.5M_\odot}\right)^{-1}
\left(\frac{\dot {M}}{\dot {M}_{\rm Edd}}\right)\,,
\end{eqnarray}
where $\dot {M}_{\rm Edd}$ is the Eddington accretion rate of the standard
neutron star with radius of 10km.
The estimated burst rate is consistent with the observed one.

Since the strange stars have thin crusts formed during
accretion from the surroundings, their maximum magnetic field
($B_{\rm max}$) is estimated through the condition$^{26}$
that in the crust the magnetic stress is equal to the maximum shear
stress, viz., $B_{\rm max}\sim (8\pi\mu\theta l/R)^{1/2}\,$,
where $l$ is the crust thickness, $\mu$ the lattice
shear modulus ($\sim 3\times 10^{26}\,{\rm dyn}\,{\rm cm}^{-2}$ at the
neutron-drip density), $\theta$ the shear angle ($\sim 10^{-2}$).
According to the crustal plate tectonics model$^{26}$, during
the spin-up of the neutron star the magnetic field lines will be pushed into
a small cap around the spin axis pole. Therefore, after the conversion
of the star the effective magnetic field ($B_{\rm eff}$)
of the strange star is calculated through the fact that
the observed magnetic moment ($\sim B_{\rm eff}R^3$) is equal to the real
one in the small cap ($\sim B_{\rm max}Rl^2$). So we have
\begin{eqnarray}
B_{\rm eff}  \sim  3\times 10^8\,{\rm G}\,\left(\frac{l}{150{\rm m}}
\right)^{5/2}\left(\frac{R}{10{\rm km}}\right)^{-5/2} \nonumber \\
\times \left(\frac{\mu}{3\times 10^{26}
{\rm dyn}\,{\rm cm}^{-2}}\right)^{1/2}\left(\frac{\theta}{10^{-2}}
\right)^{1/2}\,.
\end{eqnarray}
It is interesting to note that the magnetic-field strengths of the binary
millisecond pulsars seem to saturate at the above estimated value if the
accreted matter$^{6}$ is larger than $0.5M_{\odot}$.

It is well known that the merging of two neutron stars has been proposed
as a possible origin for cosmological $\gamma$-ray bursts$^{27}$.
Our converting model differs from the merging model as follows.
First, the merging should produce observable gravitational
waves$^{28}$. But there are no gravitational radiations in
our model if the conversion is spherically symmetric.
Future observation of gravitational waves may distinguish between
the converting and merging processes. Second, the formation rate of
compact binaries is quite uncertain, but a current estimation$^{29}$
lies in the range $10^{-5}$--$10^{-4}$yr$^{-1}$ per galaxy.
Thus the merging rate seems to be much larger than the observed burst rate.
In our scenario, the estimated rate is
consistent with the burst rate. Finally, because the strange star just
formed during the conversion has a very thin crust, the resultant fireball
is contaminated by a small amount of baryons $\le 10^{-5}M_{\odot}$.
But in the merging model the number of baryons loaded with the fireball
is unlikely to be small$^{23}$.
Therefore, the evolution of the fireball in the conversion
model is somewhat different from that in the merging model.

\begin{center}
---------------------------
\end{center}

\noindent
1. \,\,Fishman, G.J. \& Meegan, C.A. {\em Ann. Rev. Astr.
         Astrophys.}, (in press) (1995).\\
2. \,\,Mao, S. \& Paczy$\pn$ski, B. {\em Astrophys. J.} {\bf 388},
         L45-L48 (1992).\\
3. \,\,Piran, T. {\em Astrophys. J.} {\bf 389}, L45-L48 (1992).\\
4. \,\,Dermer, C.D. {\em Phys. Rev. Lett.} {\bf 68}, 1799-1802 (1992).\\
5. \,\,Hartmann, D. for a recent review in {\em The Lives of the
          Neutron Stars} (eds M.A. Alpar, $\ddot {\rm U}$. Kizilo$\check
          {\rm g}$lu \& J. van Paradijs), 495-518
          (Kluwer Academic Publishers, 1995).\\
6. \,\,van den Heuvel, E.P.J. \& Bitzaraki, O. {\em Astr. Astrophys.}
         {\bf 297}, L41-L44 (1995).\\
7. \,\,Witten, E. {\em Phys. Rev.} {\bf D30}, 272-285 (1984). \\
8. \,\,Alcock, C., Farhi, E. \& Olinto, A. {\em Astrophys. J.}
          {\bf 310}, 261-272 (1986).\\
9. \,\,Haensel, P., Zdunik, J.L. \& Shaeffer, R. {\em Astr. Astrophys.}
          {\bf 160}, 121-128 (1986).\\
10. Pines, D. \& Alpar, M.A. {\em Nature} {\bf 316}, 27-31 (1985).\\
11. Serot, B.D. \& Uechi, H. {\em Ann. Phys.} {\bf 179}, 272-293 (1987).\\
12. Baym, G. in {\em Neutron Stars: Theory and Observation}
          (eds J. Ventura and D. Pines), 21-36 (Kluwer Academic Publishers,
          1991).\\
13. Link, B., Epstein, R.I. \& Van Riper, K.A. {\em Nature}
          {\bf 359}, 616-618 (1992).\\
14. Alpar, M.A., Chau, H.F., Cheng, K.S. \& Pines, D. {\em Astrophys.
          J.} {\bf 409}, 345-359 (1993).\\
15. Alpar, M.A., Chau, H.F., Cheng, K.S. \& Pines, D. {\em Astrophys.
          J.} {\bf 427}, L29-L31 (1994).\\
16. Pandharipande, V.R., Pethick, C.J. \& Thorsson, V. {\em Phys. Rev.
          Lett.}, (in press) (1995).\\
17. Wiringa, R.B., Fiks, V. \& Fabrocini, A. {\em Phys. Rev.}
          {\bf C38}, 1010-1037 (1988).\\
18. Olinto, A. {\em Phys. Lett.} {\bf B192}, 71-75 (1987).\\
19. Horvath, J.E. \& Benvenuto, O.G. {\em Phys. Lett.} {\bf B213},
          516-520 (1988).\\
20. Dai, Z.G., Peng, Q.H. \& Lu, T. {\em Astrophys. J.} {\bf 440},
          815-820 (1995).\\
21. Glendenning, N.K. \& Weber, F. {\em Astrophys. J.} {\bf 400},
          647-658 (1992).\\
22. Haensel, P., Paczy$\pn$ski, B. \& Amsterdamski, P.
          {\em Astrophys. J.} {\bf 375}, 209-215 (1991).\\
23. M$\me$sz$\ma$ros, P. \& Rees, M. {\em Astrophys. J.}
          {\bf 397}, 570-575 (1992).\\
24. M$\me$sz$\ma$ros, P. \& Rees, M. {\em Astrophys. J.}
          {\bf 405}, 278-284 (1993).\\
25. $\ddot {\rm O}$gelman, H. in {\em Timing Neutron Stars} (eds
          H. $\ddot {\rm O}$gelman \& E.P.J. van den Heuvel), 169-189 (Kluwer
          Academic Publishers, 1989).\\
26. Ruderman, M.A. {\em Astrophys. J.} {\bf 382}, 576-586 (1991).\\
27. Eichler, D., Livio, M., Piran, T. \& Schramm, D.N. {\em Nature}
          {\bf 340}, 126-128 (1989).\\
28. Kochanek, C.S. \& Piran, T. {\em Astrophys. J.} {\bf 417},
          L17-L20 (1993).\\
29. Tutukov, A.V. \& Yungelson, L.R. {\em Mon. Not. R. astr. Soc.}
          {\bf 260}, 675-678 (1993).\\

\begin{center}
---------------------------
\end{center}

\noindent
ACKNOWLEDGEMENTS. We would like to thank M.A. Alpar, H.F. Chau, J. Fry,
S. Ichimaru, W. Kluzniak, F. Lamb, M.A. Ruderman and J.W. Truran for many
enlightening discussions. }

\end{document}